# Stability of Crossed-Field Amplifiers

Christopher Swenson, Ryan Revolinsky, Adam Brusstar, Emma Guerin, Nicholas M. Jordan, Y. Y. Lau, Ronald Gilgenbach

Plasma, Pulsed Power, and Microwave Laboratory

Department of Nuclear Engineering and Radiological Sciences

University of Michigan, Ann Arbor, MI 48109

## Abstract

This research examines the stability of crossed-field amplifiers (CFAs) and characterizes their different modes of operation: amplification, driven oscillation, and self-excited oscillation. The CFA used in this paper is the Recirculating Planar Crossed-Field Amplifier (RPCFA), which is a high power (MW) pulsed (300 ns) amplifier that operates around 3 GHz. Initially, the RPCFA is shown to be a stable amplifier with moderate gain (5.1 dB), but by either reducing the anode-cathode (AK) gap spacing or increasing the driving current, the amplifier operation transitions from amplification to oscillation. Depending on the operating conditions, these oscillations are either driven by the input RF signal or self-excited. These self-excited oscillations can have a lower synchronization phase velocity than the maximum velocity in the electron beam, implying that slower electrons within the Brillouin hub can interact with electromagnetic modes on the RF circuit. A cold tube analysis of the RPCFA shows that the Q-factor of certain modes on the RF circuit varies significantly when the AK gap geometry of the RPCFA is altered which leads to a discrete shift in operating frequency. The operation of the RPCFA close to Hull cutoff is found to share some key features of magnetically insulated transmission line oscillators (MILO) that could also explain the dramatic frequency shift. Instantaneous phase analysis by Hilbert transforms can be used, in conjunction with the frequency and output power analysis, to determine the onset of the transition from amplification to oscillation, and to characterize the oscillation.

## I. Introduction

Crossed-field amplifiers (CFAs) are broadband microwave amplifier variants of magnetrons capable of producing high-power microwaves (HPM) at relatively high efficiency [1] – [11]. Unlike magnetrons, there is a scarcity of published research on the design, operation, and limitations of CFA. To the knowledge of the authors, there is also no theory that fully encapsulates the operation of CFAs. Despite this, there exist successful simulations of CFAs and experimental findings which improve CFA performance [12] – [25]. However, the understanding of CFA operation remains primitive compared to linear beam tube devices such as traveling wave tubes and klystrons. This paper examines the operational limitations of CFA stability through simulations and experiments on the Recirculating Planar Crossed-Field Amplifier (RPCFA [15],[16],[33]), as well as using an instantaneous phase analysis technique to reveal various operational states. We also suggest a method to better elucidate the operational stability of CFAs that was not considered in previous studies of RF amplifiers.

Without an adequate theory for CFAs, complete analysis of CFA performance is challenging. The analyses in this paper assume that Brillouin flow is the prevalent equilibrium state for distributed emission CFAs [26] – [32]. The particle-in-cell (PIC) simulations of the RPCFA show an electron beam flow profile consistent with the Brillouin flow description. Using the Brillouin flow model, the unperturbed beam profile can be determined as well as the Hull-cutoff condition and Buneman-Hartree synchronous condition [31]. We assume that the input RF on the slow-wave-structure perturbs the electron beam to form spokes which grow from the Brillouin flow. The electrons in the spokes convert their potential energy to the growing RF signal as they move towards the slow-wave-structure. Thus, the maximum output power of CFAs is limited by the electron beam current in the spokes traveling along the slow-wave-structure.

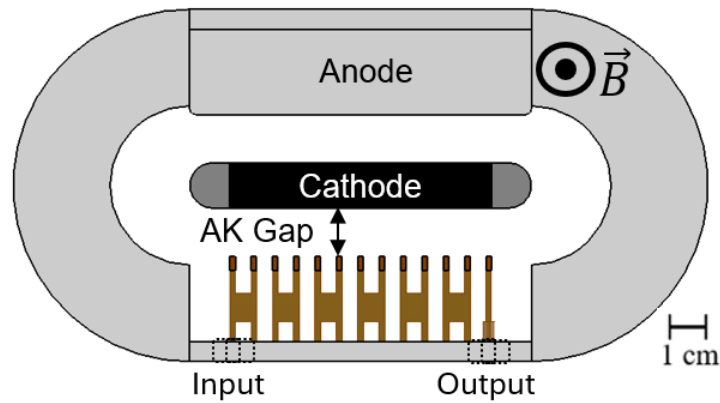

Figure 1: Cross section of the recirculating planar crossed-field amplifier

Figure 1 shows the cross-sectional view of the RPCFA examined in this paper. The RPCFA is a novel CFA design based on the Recirculating Planar Magnetron and utilizes a larger beam emission area and drift space region compared to conventional CFAs [33]. The RPCFA is a forward wave, reentrant, distributive emission CFA with coaxial ports. In previous studies [16], the RPCFA was driven with input RF sources ranging in frequency from 2.40 to 3.05 GHz and powers of 1 to 800 kW. When the injected RF power was greater than 150 kW at 3 GHz, the mean gain was 8.7 ± 0.6 dB with peak output powers reaching 6 MW before RF breakdown. The RPCFA was also shown to be zero-drive stable: with zero RF input, the RPCFA does not produce any RF output power [15], [16]. The anode-cathode (AK) gap spacing for the RPCFA experiment was set to 20 mm and the cathode used a brazed carbon fiber fabric emitter [34]. Thus, this configuration will be referred to as the nominal configuration for the RPCFA in this paper.

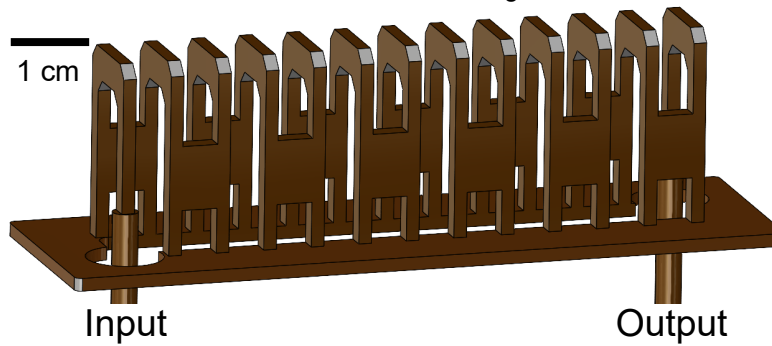

Figure 2: Meander line slow-wave-structure for the RPCFA with 12 cells.

The slow-wave-structure for the RPCFA is a meander line design shown in Fig. 2. The pitch (period) is 7.5 mm with a total cell height of 30 mm and a strap height of 10 mm. At 3 GHz, the phase velocity of the circuit is $0.3c$ with a phase advance per pitch of $\pi/2$ (forward wave). The total length of the slow-wave-structure is 12-unit cells, 90 mm, which corresponds to 3 wavelengths at the input frequency.

Section II presents the simulations of the RPCFA and examines the impact of driving current (current measured through the cathode) and AK gap spacing on the stability of the amplifier. Section III presents experimental results using parameters in the RPCFA simulations. Section IV presents the Hilbert transform, and its use for an instantaneous phase analysis of the operational state of amplifiers, in both simulations and experiments. Section V concludes.

## II. Simulations

### A. Nominal Simulations

The RPCFA represented in Fig. 3 was simulated using CST-Particle Studio's three-dimensional particle in cell (PIC) solver [14]. To mimic our pulsed power driver, simulation time was set to 300 ns with a voltage rise time of 200 ns and a steady state time of 100 ns (Fig. 4). For simulations with an input RF signal, the input signal was sinusoidal with a constant average power throughout the simulation. Voltage and current monitors are positioned at the input and output of the slow-wave-structure as well as the cathode. Electrons are emitted from the cathode with negligible momentum via the explosive emission model. The model has 1.5 million hexahedral mesh cells. The hexahedral mesh geometrically resolves most of the RPCFA well due to the rectangular geometry in the interaction region. A convergence study was performed and revealed that meshes greater than 1 million did not change the results. Instantaneous output power is measured at port 2 as seen in Fig. 3. Average power is defined in this paper as a moving average over one RF period, 0.33 ns, thus, peak instantaneous power is twice the average power.

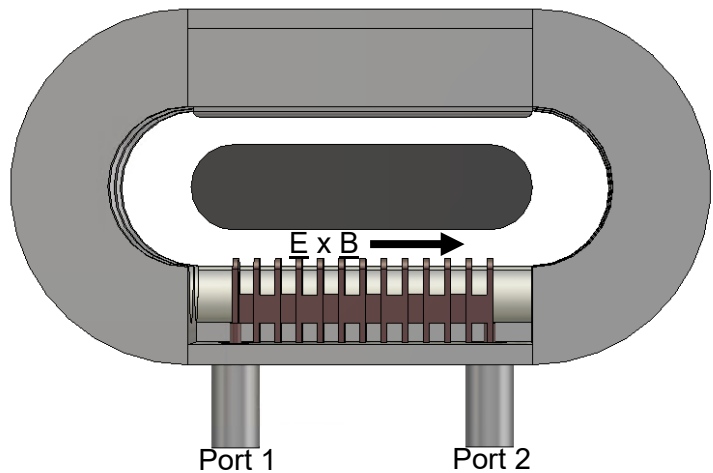

Figure 3: CST model of the RPCFA depicting the ports, electron beam direction

Nominally, the RPCFA has an anode-cathode gap of 20 mm, a cathode voltage of -330 kV, an externally applied magnetic field of 0.2 T, and slow-wave-structure pitch of 7.5 mm. With a 2 MW, 3.05 GHz input signal, the RPCFA had simulated gain of 5.1 $\pm$ 0.5 dB and exhibited relatively low current of 60 A, which is seen in Fig. 4. The reported current from these simulations is the current through the cathode which encapsulates both the (**E** X **B**) beam current and the current that moves from the cathode to the anode. The nominal design was determined by previous experiments [16] and the parameters needed for stable amplification, which will be more thoroughly explained later in this paper. This simulation agrees with the previous experimental results which showed an average peak gain of about 8 dB and thus will be used as a benchmark for the other simulations shown in this paper. These simulation results are significantly different from the MAGIC code simulation of this device [15], discussed in subsequent sections, which predicted 13.5 dB of gain and 100s A current.

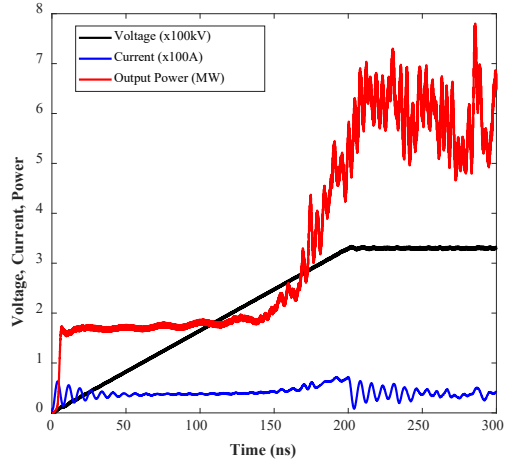

Figure 4: Average output power for a RPCFA CST Simulation with an anode-cathode gap of 20 mm and 2 MW, 3.05 GHz input signal

With zero RF input (displayed in Fig. 5), the RPCFA produces a broadband (0–6 GHz) output signal which can reach 10s of kW. During operation, a large virtual cathode, that protrudes from the cathode surface, forms near the discontinuity in the anode, which is located between the anode bend and port 1 (seen in Fig. 3). The presence of the virtual cathode perturbs the electron beam leading to electron vortices which can produce RF noise [35]. Overall, due to the relatively small output RF (10s of kW) when compared to the normal operational state (MW), the device operates as a stable amplifier, defined as zero-drive stable.

This paper examines this stability threshold and investigates different operating parameters which can lead to oscillations. The first parameter that will be explored in simulation is the operating current. Crossed-field devices (as well as other O-type tubes) can be driven to oscillate by ramping up the cathode current [3]. The current is increased by increasing the cathode potential. The current in the nominal case (Fig. 4) was noted to be very low for an operating CFA, which could impact the performance of the device.

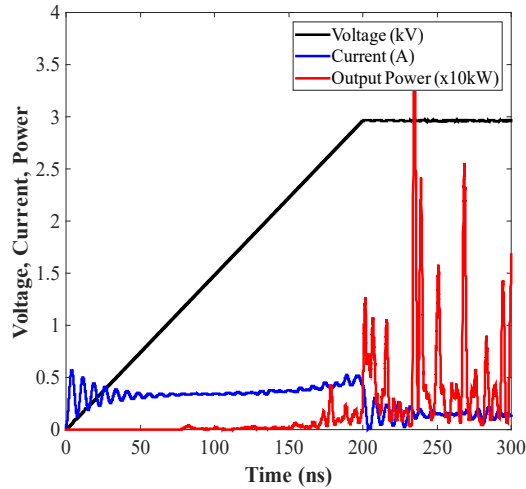

Figure 5: RPCFA CST simulation with an anode-cathode gap of 20 mm and zero RF input. The RPCFA is seen to be zero-drive stable

## B. Current-Driven Simulations

The current-driven simulations have a total simulation time of 300 ns with a rise time of 200 ns; the driving current is defined as the current from the cathode to the anode. The driving current is initially set to 0 A with a linear rise over 200 ns to the user defined input value for the steady state current. The corresponding cathode voltage for these simulations is determined by the CST solver. The driving current trace does not follow the linear rise though steady state is achieved in most simulations. Figure 6 shows the effect of increasing the driving current (measured current through the cathode) on the RPCFA where initially at 5 A, the RPCFA does not reach a high enough voltage to emit cathode electrons. In the presence of a low current electron beam (Fig 6a), the undulations in the output power grow in time due to internal reflections. When the RPCFA is operating at 60 A, for example, the average power begins to fluctuate at late time (Fig. 4), while the applied voltage remains constant. At 500 A (Fig 6b), the performance matches what has been seen in recirculating CFAs (forward or backward wave) [16], [18] where there is a distinct power peak that occurs when the voltage approaches steady state, followed by an RF power drop to a lower steady state value. This peak RF power phenomenon implies that recirculating CFAs tend to operate best with a rising voltage. This could be due to beam interactions with high Q modes outside of the steady state parameters resulting in high power oscillations which die down when the beam moves away from the high Q mode. Alternatively, the device may operate better when energy is added to the beam during operation, which prevents detuning of the beam and slow-wave structure. Understanding this phenomenon could lead to better steady state performance of recirculating CFAs and will be investigated in the future.

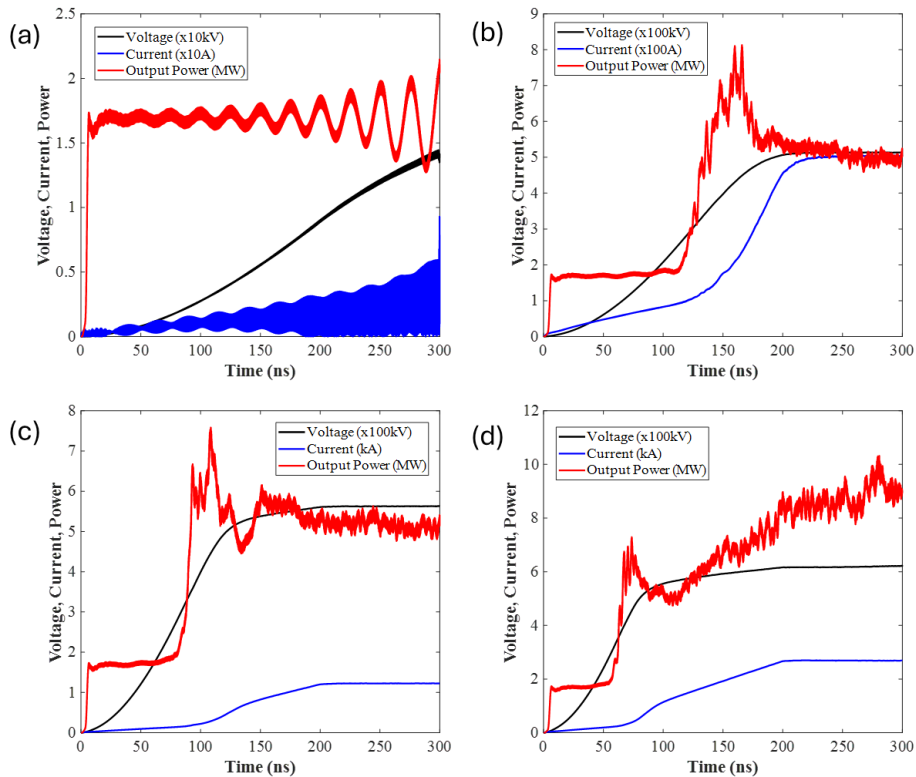

Figure 6: CST simulations for the RPCFA with a 20 mm AK gap, 0.2 T external magnetic field, and RF input of 2 MW at 3 GHz with varying driving currents:(a) 5 A (b) 500 A (c) 1.2 kA and (d) 3 kA. The voltage is determined by the CST solver given the input driving current. Despite large changes in cathode current, the RPCFA exhibits minor changes in steady state output power, indicating some operational range. Oscillations can occur when the driving current exceeds some limit.

The 1.2 kA current-driven case, Fig 6c, shows similar performance to the 500 A case with startup occurring sooner and steady state power only slightly higher. At 3 kA driving current (Fig. 6d), the RPCFA exhibits runaway output power at steady state voltage which suggests growing oscillations in the device. The large changes in device impedance observed in these RPCFA simulations mimic the large changes in critical current density which occur near the Hull cutoff for crossed-field devices [31], [36], [37], [38].

These RPCFA simulations depict three operational states which are dependent on cathode voltage/current: no amplification (Fig. 6a), amplification (Fig. 4, Fig. 6b, 6c), and oscillation (Fig. 6d). For example, the amplification regime for this RPCFA is shown to be 300 kV – 550 kV where operating below this voltage range results in no amplification and operating above this voltage range results in oscillations. The steady state output RF power within the amplification regime (300 kV – 550 kV) appears to be independent of the cathode voltage/current suggesting this RPCFA has a nonlinear voltage-gain relationship, and the device operates at a saturated limit. This phenomenon has been observed since the beginning of the CFA by Brown where he observed nonlinear gain and phase characteristics from the original platinotron [7]. However, the peak power in the amplification regime, which occurs during the voltage rise (seen in Fig. 4 and Fig. 6b, 6c), is dependent on the cathode voltage and has been observed in other CFAs [12 pp. 19,20,58], [18]. This implies that CFAs operate differently during the voltage rise when compared to steady state.

Examination of frequency spectra is one way to determine if the device is oscillating and can be used to discern the oscillation mode. Figure 7a presents output spectra of the RPCFA with 5 A of driving current versus 3 kA (Fig. 7b). At 5 A driving current, the electron beam should minimally detune the RF signal, thus the signal should ideally be a single tone; however, significant side bands are observed. Although the device is clearly not oscillating, these data along with Fig 6a., suggest that the RPCFA exhibits internal reflection effects. These internal reflections could lead to RF phase distortion at the output and, thus, side bands in the spectra.

Figure 7b shows the output frequency spectra of the 3 kA driving current. There exist two distinct peaks: the first peak is at the input frequency (3 GHz) and the second peak is at a much higher frequency (8.7 GHz). This could suggest that the RPCFA operates as a driven oscillator at 3GHz and/or a self-excited oscillator at 8.7 GHz, simple frequency spectra examination is not definitive concerning device oscillation. To determine whether a CFA operates as an amplifier or a driven oscillator requires a novel analysis method which is employed and described later in this paper.

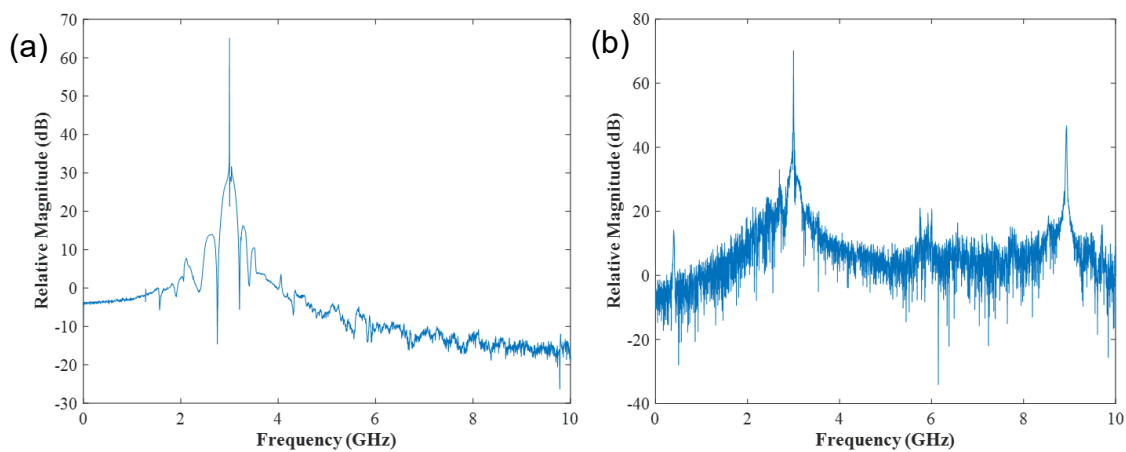

Figure 7: Output frequency spectra in relative amplitude with RF input of 2 MW at 3 GHz, plotted for: (a) 5 A current drive case and (b) 3 kA current drive case. The 5 A case shows a mostly single frequency spectrum but with significant side bands. At 3 kA drive current, the noise floor of the signal increases significantly and a higher frequency signal is observed. The AK gap spacing is 20 mm.

## C. Changing AK Gap Simulations

The current-driven simulations showed that, given high enough voltage (> 500 kV), the device oscillates and the RPCFA might operate as a driven oscillator rather than an amplifier. This effect is difficult to prove experimentally (needs a very high voltage source) and can be avoided by simply not "over-volting" the amplifier. Another method for mimicking an increase in applied voltage (increasing the electric field) is by reducing the anode-cathode (AK) gap. By reducing this gap and keeping the nominal voltage (-300 kV), the impact of diode impedance, capacitance, and applied electric field can be studied in simulation and experiment. The AK gap was reduced in two different ways, with the same effect: 1) increasing the cathode thickness or 2) placing the cathode closer to the slow-wave structure.

Figure 8 shows the output power of the RPCFA with zero input RF, -300 kV cathode potential and a 0.2 T external field. The black line is the transition to steady state voltage at 200 ns. The current is listed in Table 1. From Fig. 5, with a 20 mm or 17.5 mm AK gap, the RPCFA was shown to be zero-drive stable with an applied voltage of -300 kV. However, when the AK gap is reduced below 17.5 mm, the RPCFA oscillates with zero input signal. When the AK gap is reduced below 12.5 mm, the device operates near Hull cutoff and the output power increases exponentially.

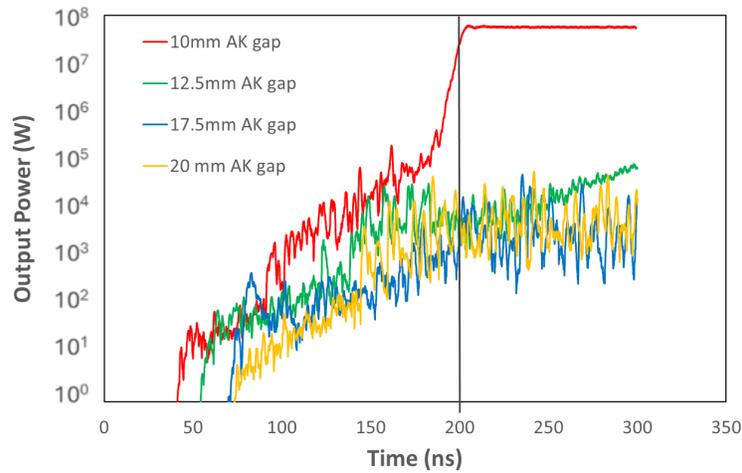

Figure 8: Output power of the RPCFA with zero RF input, cathode voltage of -300 kV, and external magnetic field on 0.2 T. At 20 mm AK gap, the device was shown to be zero-drive stable; however, by decreasing the gap below 17.5 mm or less, the device oscillates.

Table 1 shows the current, Hull cutoff magnetic field, electric field, and magnetic insulation ratio for each gap spacing *without* input RF. The device operates below the Hull cutoff magnetic field ($B_H$) when the AK gap is 10 mm. The large change in current when the AK gap spacing is reduced is due to the large changes in device impedance which occur near Hull cutoff [31], [36], [37], [38]. Table 2 is the tabulated data from the simulations depicted in Fig. 6, which had input RF of 2 MW.

In either case, continuous output power growth was observed at steady state cathode voltage and current when the degree of magnetic insulation was 1.18. With RF input, Fig. 6d, the current was 3 kA when this phenomenon occurred. Figure 8 (*without* RF input) shows that same phenomenon, in which output power continues to increase at the steady state voltage but at a much lower current (230 A, Table 1); *thus, the current alone cannot be used to determine oscillation thresholds in the RPCFA.* Since the device oscillated without input RF, the oscillations in the RPCFA can be self-excited.

| AK GAP (MM) | CURRENT (A) | HULL CUTOFF FIELD (T) | ELECTRIC FIELD (MV/M) | MAGNETIC INSULATION RATIO ($B/B_H$) |
|---|---|---|---|---|
| **20** | 15 | 0.10 | 15 | 2.00 |
| **17.5** | 20 | 0.12 | 17 | 1.67 |
| **12.5** | 230 | 0.17 | 24 | 1.18 |
| **10** | 2800 | 0.21 | 30 | 0.95 |

Table 1: Tabulated data from the simulations seen in Fig. 8 where the cathode potential was held constant while the AK gap spacing was changed. Cathode current, Hull cutoff magnetic field $B_H$, DC electric field, and magnetic insulation ratio for each simulated anode-cathode gap seen in Fig. 8 with zero input RF. Cathode voltage was set to -300 kV. Externally applied field, B, is 0.2 T, thus the 10 mm gap case is under insulated (B < $B_H$).

| CURRENT (A) | VOLTAGE (KV) | HULL CUTOFF FIELD (T) | ELECTRIC FIELD (MV/M) | MAGNETIC INSULATION RATIO ($B/B_H$) |
|---|---|---|---|---|
| **5** | 30 | 0.03 | 1.5 | 6.67 |
| **500** | 500 | 0.14 | 25 | 1.43 |
| **1200** | 580 | 0.16 | 29 | 1.25 |
| **3000** | 610 | 0.17 | 30.5 | 1.18 |

Table 2: Tabulated data from the simulations seen in Fig. 6, where the AK gap spacing was 20 mm, the RF input was 2 MW, the externally applied magnetic field was 0.2 T, and cathode current was varied. Note that the 3000 A case is rather close to Hull cutoff ($B/B_H$ = 1).

From this study, the difference between the CST-PIC simulations and the previous MAGIC simulations is identified. Though the previous RPCFA experiments had an AK gap of 20 mm, the original MAGIC simulations were performed with an AK gap spacing of 15 mm. The CST simulation given here cannot be adequately compared with the previous MAGIC simulation [15], due to the demonstrated sensitivity to the circuit parameters which were insufficiently specified in the previous simulations, and to the lack of a benchmark of these two codes on crossed-field devices.

Figure 9 depicts the simulated electron beam during steady state for the 10-mm AK gap case with zero RF input. CST predicts an average of 60 MW RF power when the AK gap is reduced to 10 mm, but the device is not magnetically insulated (B < $B_H$), as shown in Table 1. This RPCFA simulation is akin to simulating a MILO whose magnetic field is provided externally, with the MILO characteristic [38] that it operates very close to Hull cutoff, in the π-mode with spokes generated only in the downstream region. In experiments, the device would electrically short to the slow-wave structure and cut off any RF generation. Figure 8 and Table 1 illustrate, perhaps for the first time, a definitive trend that shows oscillation power increasing as a CFA approaches Hull cutoff. However, without simulating ablation and plasma generation in the gap and around the slow-wave structure, CST could not accurately predict operation near Hull cutoff for crossed-field devices in experiments.

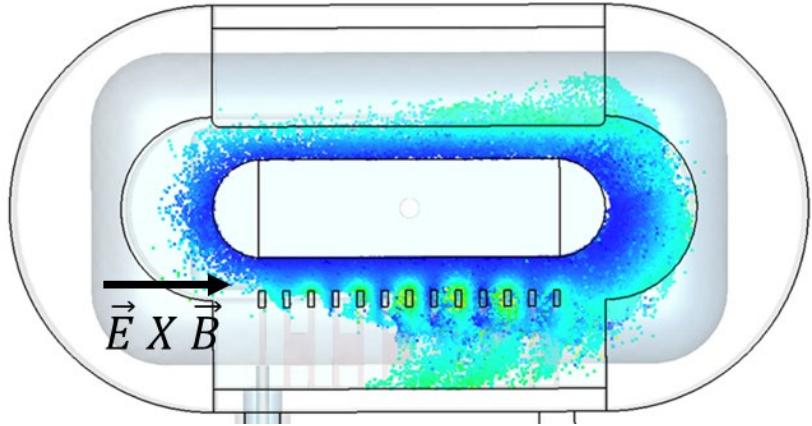

Figure 9: CST simulation of the RPCFA with an AK gap of 10 mm and no RF input, applied cathode voltage of -300 kV, and external field of 0.2 T. The device still oscillates in simulation even though the RPCFA is not magnetically insulated (B < $B_H$).

Frequency spectra of the output RF signal in the RPCFA simulations with zero RF input for the two different AK gap spacings are presented in Fig. 10. The 16 mm AK gap was chosen because this is the largest gap where a single frequency peak is present in the spectra. Larger AK gaps (e.g., 17.5 mm) which produce output power have a noisy spectrum with no clear dominant mode. The 3 GHz oscillation frequency is the designed operational frequency of the RPCFA. As seen in the current-driven simulations, this RPCFA also oscillates at 3 GHz but with zero input. This result implies that the device can act as a self-excited oscillator at 3 GHz. This makes distinguishing the RPCFA operation difficult, since it can act as one of three types of devices all at the same frequency: 1) amplifier, 2) driven oscillator, and 3) self-excited oscillator [42]. If the amplifier oscillates at the driving/input frequency, frequency spectra alone will not be sufficient to determine whether the amplifier is oscillating. A method for improved discrimination between oscillating and amplifying operation will be discussed in Section IV, where the phase analysis will be performed on both simulation and experimental results.

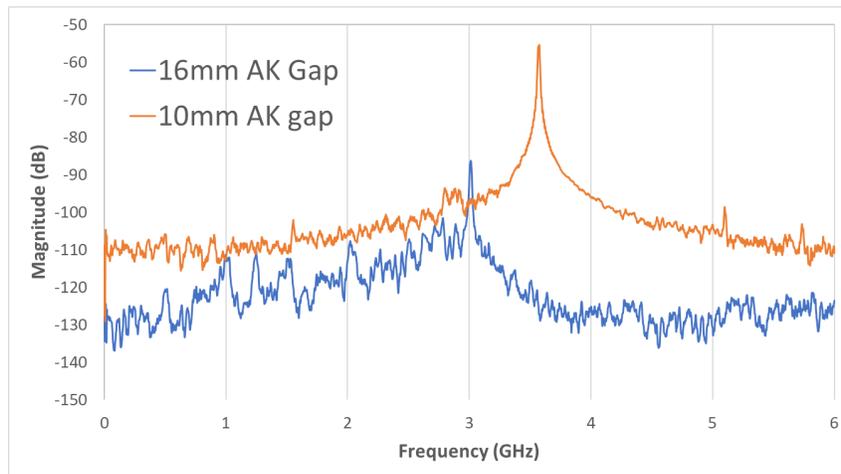

Figure 10: Frequency spectra of RPCFA simulations with zero RF input, cathode voltage of -300 kV, and external magnetic field of 0.2 T for AK gaps of 16 mm and 10 mm.

When the AK gap spacing is reduced to 10 mm, the RPCFA switches to a different mode, causing the zero-drive oscillations to shift up to 3.5 GHz as seen in Fig. 10; reduction of the AK gap spacing is the cause of this shift. In the current-driven simulations, higher frequency modes were observed but the original 3 GHz mode was still present (Fig. 7b).

To explain the observed frequency up-shift when the AK gap spacing is reduced, the dispersion relation for the RPCFA with 20 mm gap spacing is plotted in Fig. 11. The infinite unit cell dispersion relation is represented by the black curve and the red dots along the curve are the discrete modes that exist in the finite model (note that the RPCFA has 12 unit cells). The other lines represent the beam modes at various electron beam velocities. From this dispersion relation, the 3 GHz mode is seen to be a forward wave mode that exists at π/2 phase advance and the 3.5 GHz mode is seen to be the π mode. The 0.3c electron beam (blue) line represents the electron velocity at the top of the Brillouin hub and is the designed beam velocity. This 0.3c line intersects with the 3 GHz π/2-mode as intended. Given a 20 mm AK gap spacing, a 0.175c beam velocity would interact with the 3.5 GHz mode; i.e., a slower beam is required to synchronize with the π mode. However, when the AK gap spacing is reduced to 10 mm, the average beam velocity is increased, and the top of the Brillouin hub exceeds 0.4c. A higher beam velocity would imply that the beam should interact with a lower frequency mode rather than a higher frequency mode. The changing AK gap spacing does not change the dispersion relation enough to explain this shift; as the AK gap spacing is reduced, all modes other than π-mode exhibit a slight reduction in frequency (<15 MHz/mm). The π-mode frequency does not change with the AK gap spacing.

One reason for the observed frequency up-shift, despite the increasing average beam velocity, is the impact that changing the AK gap spacing has on the Q factor of each mode. Initially, at 20 mm, the Q factors for the π/2-mode and π-mode are 30 and 62 respectively. When the AK gap spacing is reduced, the Q factors are altered to 16 for the π/2-mode and to 2000 for the π-mode. Thus, as the AK gap spacing is reduced, the π/2-mode's Q factor is reduced slightly, however, the π-mode's Q factor increases over an order of magnitude. This alone does not explain why there is a frequency upshift, but since a 0.175c electron beam velocity exists in the Brillouin hub, it could interact with a very high Q in the π-mode.

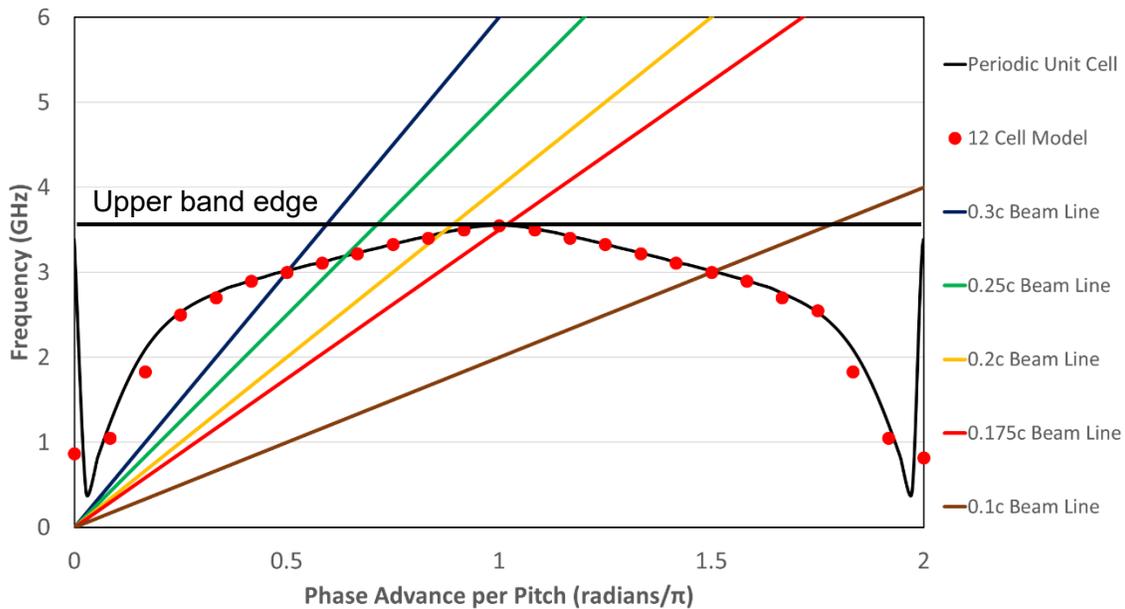

Figure 11: Dispersion relation of the RPCFA where the black curve is the infinite unit cell, the red dots are the finite modes, and the straight lines show the beam modes at various beam velocities. The AK gap spacing is 20 mm.

The electron beam in the planar RPCFA can be described by Brillouin flow, for which the Buneman-Hartree condition stipulates that the phase velocity of the operating mode synchronizes with Brillouin flow velocity at the top of the Brillouin hub (the highest velocity), and this condition for startup is proven to be the same as that obtained from the single-particle, cycloidal orbit model [31]. The distribution of electron velocities contains all velocities between 0 (at the cathode) and the maximum velocity at the top of the hub [31]. Thus, the electron layer within the Brillouin hub with the streaming velocity 0.175c may resonantly excite the high Q mode at 3.5 GHz (Fig. 11).

Note that the π-mode at 3.5 GHz is recognized as an upper band edge in the dispersion diagram, Fig. 11. For forward wave traveling wave tube amplifiers, the upper band edge is known to post a very serious threat because it has a much lower threshold current to excite absolute instability than the lower band edge [39]. If band edge oscillation proves to be a threat to a distributed emission CFA, its stability could be improved if the CFA operated in the backward wave branch, beyond the π-mode in Fig. 11.

Yet another plausible explanation for the self-excited oscillation at 3.5 GHz when the AK gap spacing is reduced to 10 mm is that such a CFA, without an RF input, operates close to Hull cutoff, like a MILO [38] (though the magnetic field is provided externally here). The appearance of the π-mode is also similar to a MILO in operation, as we pointed out above when Fig. 9 is discussed. Note that we have not identified which mechanism is responsible for this 3.5 GHz self-excited oscillation: resonant excitation of the high Q mode by the co-moving electrons in the Brillouin hub, upper band edge oscillation at the π-mode, or a MILO-like oscillation (also at the π-mode) close to Hull cutoff as the AK gap spacing is reduced.

### III. Experiments

#### A. Nominal Experiments

Experimentally, the RPCFA depicted in Fig. 1 was driven by the Michigan Electron Long Beam Accelerator with Ceramic insulator (MELBA-C) which delivers a 500 ns, -300 kV pulse with a 100-200 ns rise time [40]. The RPCFA is in a vacuum chamber which is pumped down to the $10^{-6}$ torr scale. The RPCFA slow wave structure input and output coaxial lines are connected to WR284 waveguide in vacuum through a vertical field coupler which has an upper and lower frequency cutoff of 2.4 GHz and 3.2 GHz, respectively. Two axial coils in a Helmholtz configuration are set up around the vacuum chamber to deliver a spatially uniform external magnetic field of 0.1 – 0.3 T with a long pulse (~20 ms) compared to the MELBA pulse. Two WR284 directional couplers are used to sample the input, output, and reflected RF signals. These signals are then attenuated, split and transmitted to a 6 GHz oscilloscope for frequency analysis and to microwave diodes for power analysis. Output RF waveguide is terminated in a matched load with an S11 of -19.8 dB at 3.0 GHz. An E2V MG5193 magnetron generates the 3 GHz, ~500 kW RF input for the experiment; this magnetron is protected from reflections with a waveguide isolator. AK gap spacing can be adjusted by offsetting the cathode relative to the slow-wave structure; the range of this adjustment was 12.5 mm to 20 mm. An experimental analysis of the 10 mm gap could not be done because the RPCFA arced with such a small gap spacing and damaged the SWS.

The first experimental series (Fig. 12) was performed with an AK gap spacing of 20 mm to replicate the CST simulation performed earlier (see Fig. 4). The mean peak gain of this series was 4.2 ± 3.4 dB and the peak output power was 4.2 ± 0.2 MW. For the experimental data presented, gain is calculated with the total output power (all frequencies). By removing the shots that did not produce any measurable gain, the mean peak gain increases to 5.6 ± 3 dB which is close to the simulated gain of 5.1 ± 0.5 dB. The reason peak gain is reported here instead of average gain is due to experimental limitations. As later plots show, the RF pulse shows signs of RF breakdown or plasma diode closure which prematurely terminates the RF pulse. Comparing these data to the nominal CST simulation, there is reasonable agreement with the simulated gain of 5.1 ± 0.5 dB. The major difference between the simulation and the experiment originates in the measured current value. The experimental current was 2.5 ± 0.5 kA and is not correlated with the gain, while the simulated current was only 0.06 kA. There are several reasons for this discrepancy, some of which have been stated previously. Despite this discrepancy, we will show below in Sections IV.B and IV.C that, with the use of Hilbert transform, the phase analysis reveals some common features shared by the simulations and the experiments.

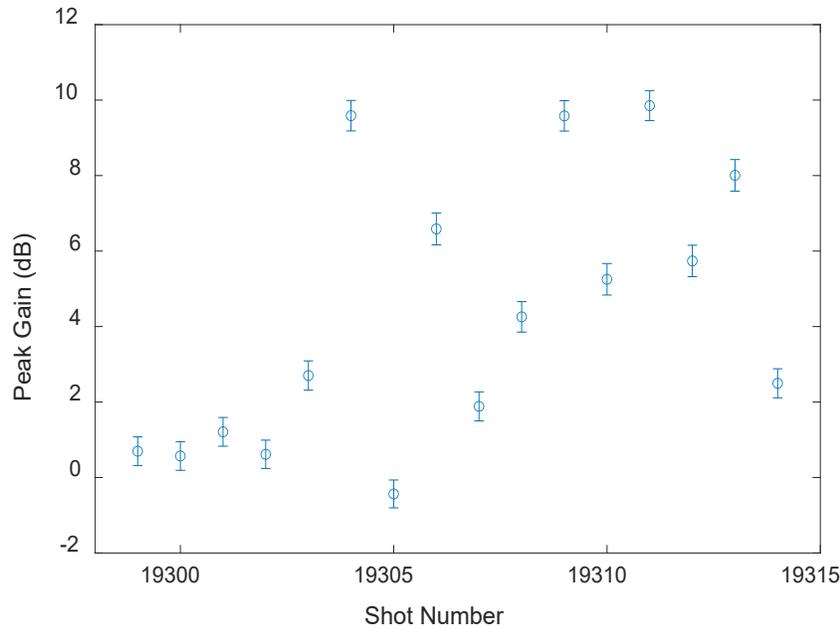

Figure 12: Experimental RPCFA peak gain with AK gap of 20 mm and average input power of 420 kW at 3 GHz. This is the nominal configuration that was simulated in CST.

The CST simulation lacks an ablation model and any plasma in the interaction region or on the slow-wave structure. Plasma generation will occur in the experiment from back bombardment of the electron beam onto the cathode which can lead to plasma sheath around the cathode effectively decreasing the AK gap spacing. Electron bombardment on the slow-wave structure can lead to significant ablation of the copper structure and plasma formation around the slow-wave structure which can lead to RF reflections. The plasma forming during the experiment can also lead to plasma diode closure where the effective AK gap spacing is reduced during the pulse. Finally, the CST simulation has a set emission model that can only emit on select cathode surfaces. This is done to reduce computational time and computational requirements. In the experiment, electrons are emitted from the entire cathode, including the end hats and stalk. Considering all these effects, the experiment is expected to have a larger current and for this current to be mostly independent of the RF gain (the simulation predicts 60 amps at high power compared to the baseline 2 kA of current in the experiment).

Figure 13 depicts some example shots from the RPCFA experiment and the electrical pulse characteristics of MELBA. In these shots, there are several important differences when compared to the CST simulations. The RPCFA (with or without RF input) does not operate on the voltage rise and the peak power of the RPCFA usually occurs towards the end of the voltage pulse after a significant amount of time at steady state voltage and current. After the peak power, the RF signal quickly goes to zero which could imply RF breakdown or plasma diode closure which are not seen in simulation. Detailed examination of Fig. 13a shows that there are two distinct operating regimes: a region with a slow increase in output power followed by a dramatic spike in output power. Figure 13c is the corresponding frequency spectrum of shot 19053, which had RF input, and shows the dominant 3 GHz signal with large side bands. This spectrum is similar to the spectrum seen in Fig. 7a which was a simulation of the RPCFA with RF input and only 5 A of driving current. This suggests that the distortions seen in this spectrum are a result of the cold characteristics of the RPCFA (internal and external reflections). From this spectrum and the zero-drive stability of this geometry, shot 19053 appears to be amplifying. This information, however, is insufficient to distinguish between amplification and driven oscillation. A further examination of this shot will be presented in the phase analysis section, Section IV.C.

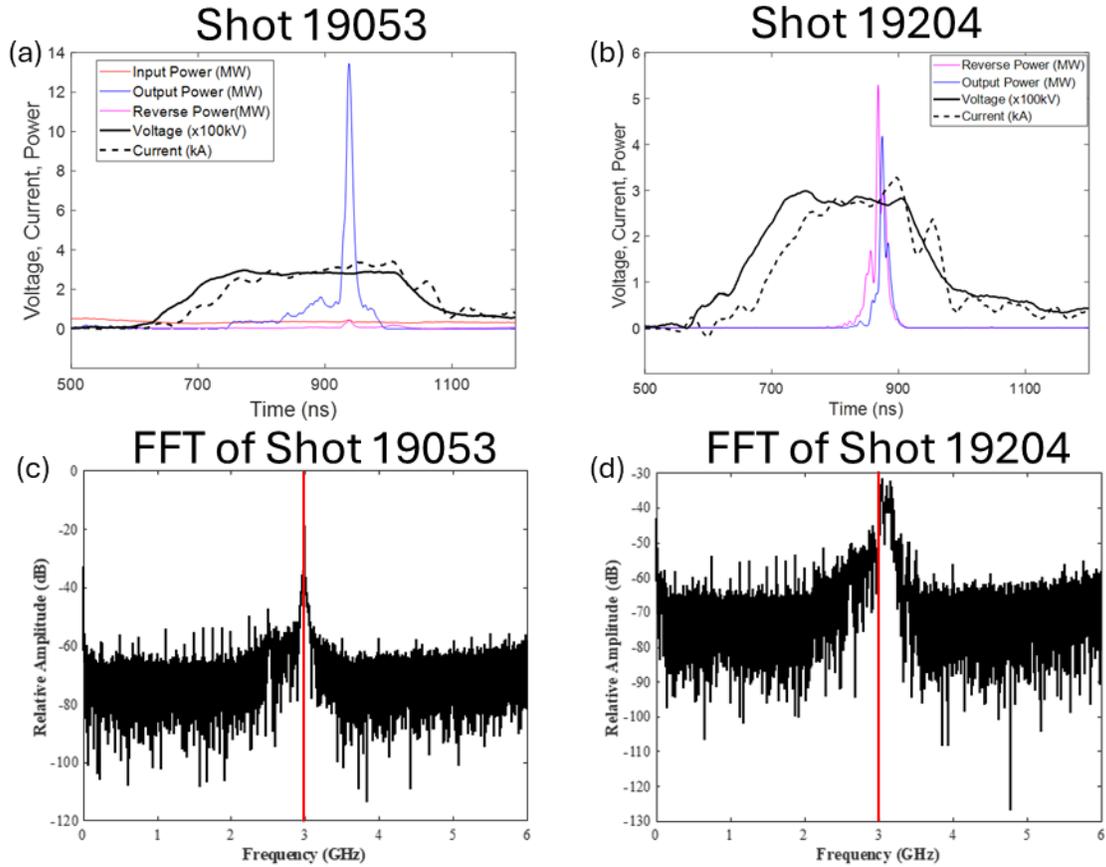

Figure 13: A pair of RPCFA experimental shots showing the applied voltage, current, and power. a) 430 kW RF input at 3.0 GHz and an AK gap spacing of 18mm (#19053); (b) Zero RF input and an AK gap spacing of 12.5 mm (#19204). Corresponding frequency spectra are shown in (c) for shot 19053 and (d) for shot 19204 with a vertical line at 3 GHz.

Focusing on Fig. 13b, a large spike in reverse power that is on the order on the forward power is observed in all zero-drive shots with high output powers (large oscillations). This reverse power spike is not seen in most shots with RF input. This reverse power could be a sign of a standing wave on the slow-wave structure that is interacting with the electron beam and growing. The corresponding frequency spectrum in Fig.13d shows a very broad spectrum which ranges from 3-3.2 GHz (there could exist a higher frequency signal that is cut off by the waveguide couplers). This frequency up-shift was predicted by simulation and results from the Q factor of the π-mode increasing with the decreasing AK gap spacing. This frequency up-shift is also experimentally observed in the 12.5 mm AK gap spacing experiments with RF input. The broad spectrum suggests that there is significant mode competition in this device when the RPCFA oscillates.

Therefore, this device seems to exhibit several different modes of operation: amplifying, driven oscillation, and self-excited oscillation. Self-excited oscillation can be induced by increased feedback, increased electric field, or by having the beam interact with a high Q factor mode. Driven oscillations have similar requirements to self-excited oscillations but require an input driving signal that primes the oscillations.

**B. AK Gap Spacing Experiments**

The following experimental series followed the same procedure as the previous one, but the AK gap spacing was reduced each time. Fig. 14 shows the aggregated results of the experimental series with average RF input power of 400 kW. These data show no correlation between the AK gap spacing and the average gain of each series. There was also no correlation between the gain and the current during any of the experiments and the current did not significantly increase with the decrease in the AK gap spacing (an average of 2.6 ± 0.4 kA at the 12.5 mm gap). The gain/power limitation of this device is a multifaceted problem where the hard limit is due to RF breakdown threshold and desynchronization of the electron beam and the RF wave on the slow-wave-structure. Desynchronization can occur due to plasma formation or the electron beam slowing down. At 20 MW, the electric field between the slow-wave structure vanes reaches about 10 kV/cm and could cause breakdown in the presence of beam electrons.

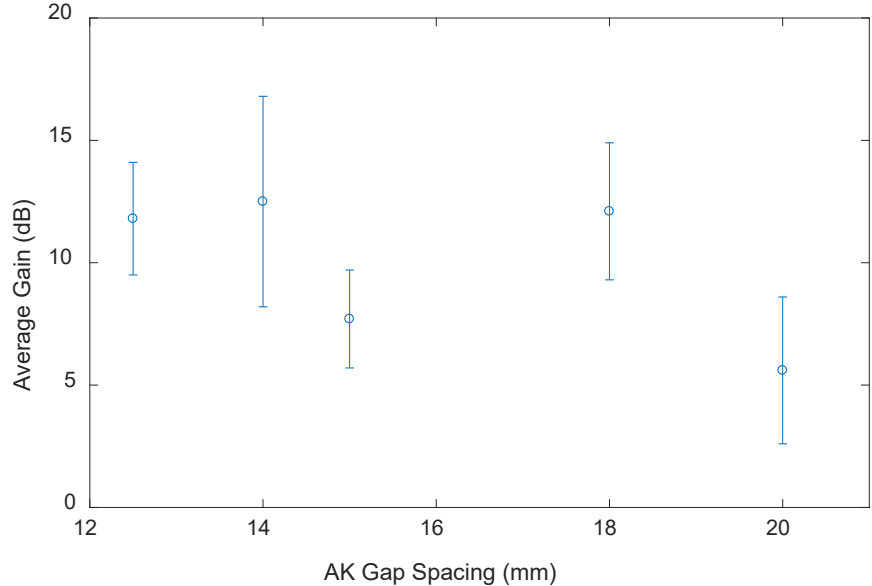

Figure 14: Average RPCFA gain over an experimental series for several different AK gap spacings, average RF input power of 400 kW

The next experimental series examined the impact of reducing the AK gap spacing without RF input. As seen in Fig. 15, the device showed little change in the average output power (each plot point represents 30 shots) measured until the gap was reduced to 12.5 mm. Thus, the RPCFA acts as an infinite gain amplifier/oscillator. Along with the dramatic increase in output power, an up-shift in the frequency spectrum is observed (Fig. 16). The significant increase in output power at 12.5 mm observed in experiment is very similar to the simulation results for 10 mm that were shown in Fig. 8. The latter simulation results, together with Fig. 9, suggested that the zero RF input RPCFA at 10 mm operates somewhat like a MILO oscillator. Indeed, from the last two rows of Table 1, the value of B/BH is very close to unity at both 12.5 mm and 10 mm gap separations This slight difference in the threshold spacing between experiment and simulation is most likely due to the incomplete modeling of various plasma formation and breakdown processes in the CST simulation. In simulation at a 10 mm AK gap spacing, the peak RF electric field between the vanes of the slow-wave structure reached over 50 kV/cm and would likely break down experimentally. Because they do not contain an emission/plasma model for the slow-wave structure, CST simulations can achieve high vane-to-vane RF electric field strengths without exhibiting RF breakdown; if no electrons are emitted from the vanes of the slow-wave structure then there cannot be breakdown. As stated previously, the emission area is limited to designated surfaces of the cathode due to computational limitations. Outgassing is also absent in simulations. Effective changes in gap spacing due to such plasma effects are thus absent in CST.

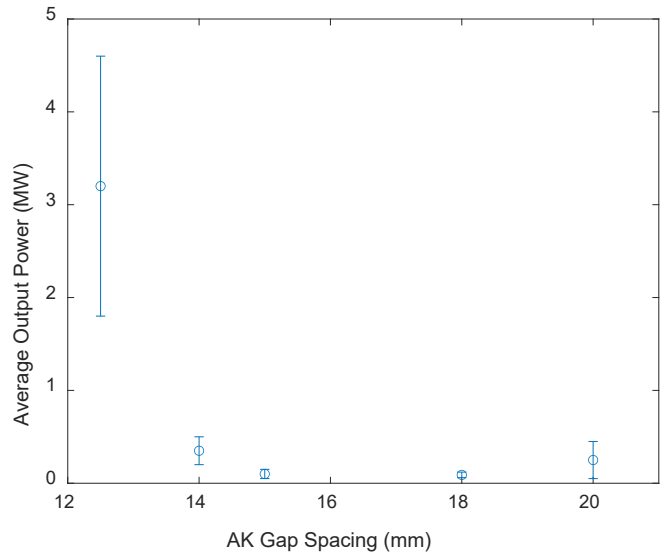

Figure 15: Experimental peak output power at several AK gap spacings with zero RF input. Each data point represents an average over 20-40 individual shots.

For AK gap spacings from 14 mm to 20 mm, the frequency spectrum is centered around 2.97 GHz, but when this gap spacing is reduced to 12.5 mm the spectrum is centered around 3.16 GHz (Fig. 16). Spectra produced from simulations (Fig. 10) are much cleaner and show single peaks while the experimental spectra are noisy and broad. The first reason for this difference is that the experiment has increased mode competition compared to simulation, which is due to the imperfections within the experimental RPCFA (electrical connections, geometry, plasma formation, etc.). The second reason is that the RF pulse duration is much shorter in the experiment which reduces the chance that a single dominant mode arises. Nonetheless, the frequency up-shift is observed and corroborates the simulations which imply that geometric cathode adjustments impact RF modes differently and a small change in the AK gap spacing can lead to significant oscillations and frequency changes. The high sensitivity to small changes in magnetic insulation near Hull cutoff is another property that is seen in MILOs [31], [38] and could suggest a fundamental transition in operation for this CFA.

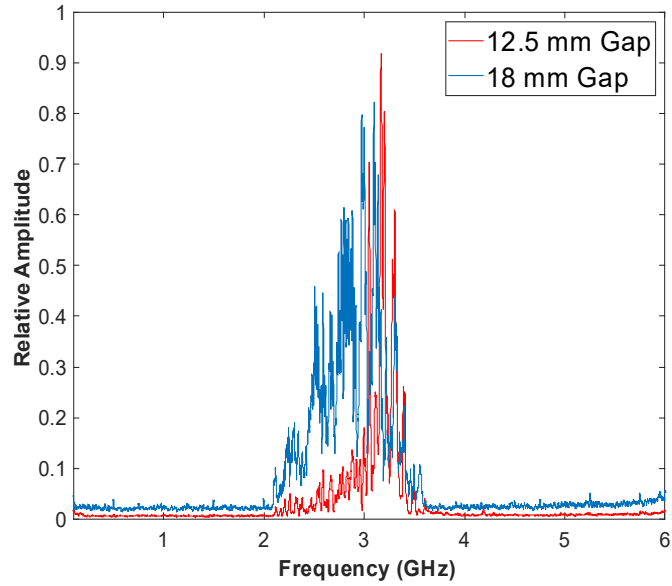

Figure 16: Frequency spectrum showing the relative amplitude of each frequency component within the output signal of shot 19019 with an 18 mm AK gap spacing and shot 19195 with a 12.5 mm AK gap spacing both without RF input

## IV. Phase Analysis

### A. Hilbert Transform

Analyzing the power trace and frequency spectrum alone is not enough to determine if the RPCFA is acting as a driven oscillator operating at the frequency of the input signal. By examining the phase difference between the input and output signal, the transition to oscillation (regardless of the type of oscillation) can be observed. The Hilbert transform is the principal value (P. V.) of the convolution of the RF voltage signal (represented by x(t)) and the simple function h(t) = 1/πt [41],

$$H(x(t)) = P.V. \frac{1}{\pi} \int_{-\infty}^{\infty} \frac{x(t)}{\tau - t} dt \qquad (1)$$

Assuming the voltage signals have the form

$$x_i(t) = A(t) \cos\left(2\pi \int_0^t f_i(t)\, dt + \theta_i(t)\right) \qquad (2a)$$

$$x_o(t) = B(t) \cos\left(2\pi \int_0^t f_o(t)\, dt + \theta_o(t)\right) \qquad (2b)$$

where the indices i and o represent the input and output signal respectively with amplitudes A and B, frequency f, and phase θ. By using the Hilbert transform on these signals, the total phase difference between the input and output can be written as

$$\Delta\theta(t) = \text{atan}\left(\frac{H(x_i(t))}{x_i(t)}\right) - \text{atan}\left(\frac{H(x_o(t))}{x_o(t)}\right). \qquad (3)$$

If we assume that the amplitudes of the signals do not change too quickly (do not change significantly over 1 RF period) and $f_{i,o}$ is sufficiently large compared to the time rate of change of $\theta_{i,o}$, substituting Eqs. (1) and (2) into Eq. (3) and simplifying yields

$$\Delta\theta(t) = 2\pi \int_0^t \Delta f(t)\, dt + \theta_o(t) - \theta_i(t) \qquad (4)$$

where $\Delta f(t) = f_o(t) - f_i(t)$. Equation (4) accounts for the phase change due to the difference in signal frequencies, the remainder may be categorically considered as phase noise. In an example given in the next paragraph, we illustrate how to separate the contributions from f and θ. Equation (4) can be simplified further if the input signal does not change; $\theta_i(t)$ and $f_i(t)$ are constants which is the case for an amplifier with constant input RF. Further, if there is not a frequency change, the phase difference measures the phase fluctuations. Therefore, the phase difference is constant in an ideal amplifier where the frequency does not change. Deviations from the ideal case where the phase difference is constant are from the output signal: $f_o$ is constant but different from $f_i$, $f_o$ is time varying, $\theta_o$ is time varying.

First, we will examine the effect of each of these changes in the output signal independently. If $f_o$ is constant but different from $f_i$ with no change in $\theta_o$, the phase difference as a function of time will be a straight line with slope $f_o-f_i$. If $f_o$ is time varying but with no change in $\theta_o$, the phase difference will be a curve which is either: smooth ($f_o(t) - f_i$ is differentiable) indicating a smooth push or pull in the output frequency; or sharp indicating new discrete frequency components that quickly enter or leave the harmonic content of the output. Finally, if $\theta_o$ is time varying, this will look similar to the case where $f_o$ is time varying because frequency is the derivative of phase; therefore, time varying phase and time varying frequency effects on the phase difference might look identical. In experiments, the phase difference will be some linear combination of all these elements and will have to be separated and identified. We may use the rightmost figure in Fig. 17a given below, which is a plot of the LHS of Eq. 4, to illustrate. The smoothed red curve represents the contribution from $\Delta f(t)$ in the RHS of Eq. 4, whereas the remaining, rapidly fluctuating part accounts for the contributions from $\theta_o(t) - \theta_i(t)$.

Thus, by using Eq. 4, an amplifier can be rigorously examined, and the onset of oscillation can be determined. If the device does transition to a driven oscillator, the original phase relationship will change abruptly as the operating state changes. If there is self-excitation, then there will be no relationship between the input and output phase until a single mode dominates. In summation, sudden large changes in the phase difference are a result of a transition of operating states, i.e., amplification to oscillation.

### B. Simulated Phase Analysis

Figure 17 shows different simulated modes of operation and the corresponding frequency spectra and phase difference graphs. All simulations shown in Fig. 17 have a voltage rise time of 200 ns to 300 kV. Figure 17a depicts the nominal operation where there is low gain (like Fig. 4), and the frequency spectrum is monochromatic which suggests amplification. The phase difference graph has five distinct features. The initial spike is from the delay between the input and output signal at the start of the simulation which transitions to a flat line when the RF signal reaches the output port. Then, a constant phase difference (10 ns – 60 ns) when there is no electron beam present. Next, there is an observable (linear) increase in phase (90 ns – 120 ns) which coincides with the increasing beam current as the voltage rises. Examining the linear region (90 ns – 120 ns), a frequency difference of ~3 MHz is calculated using Eq. 4.

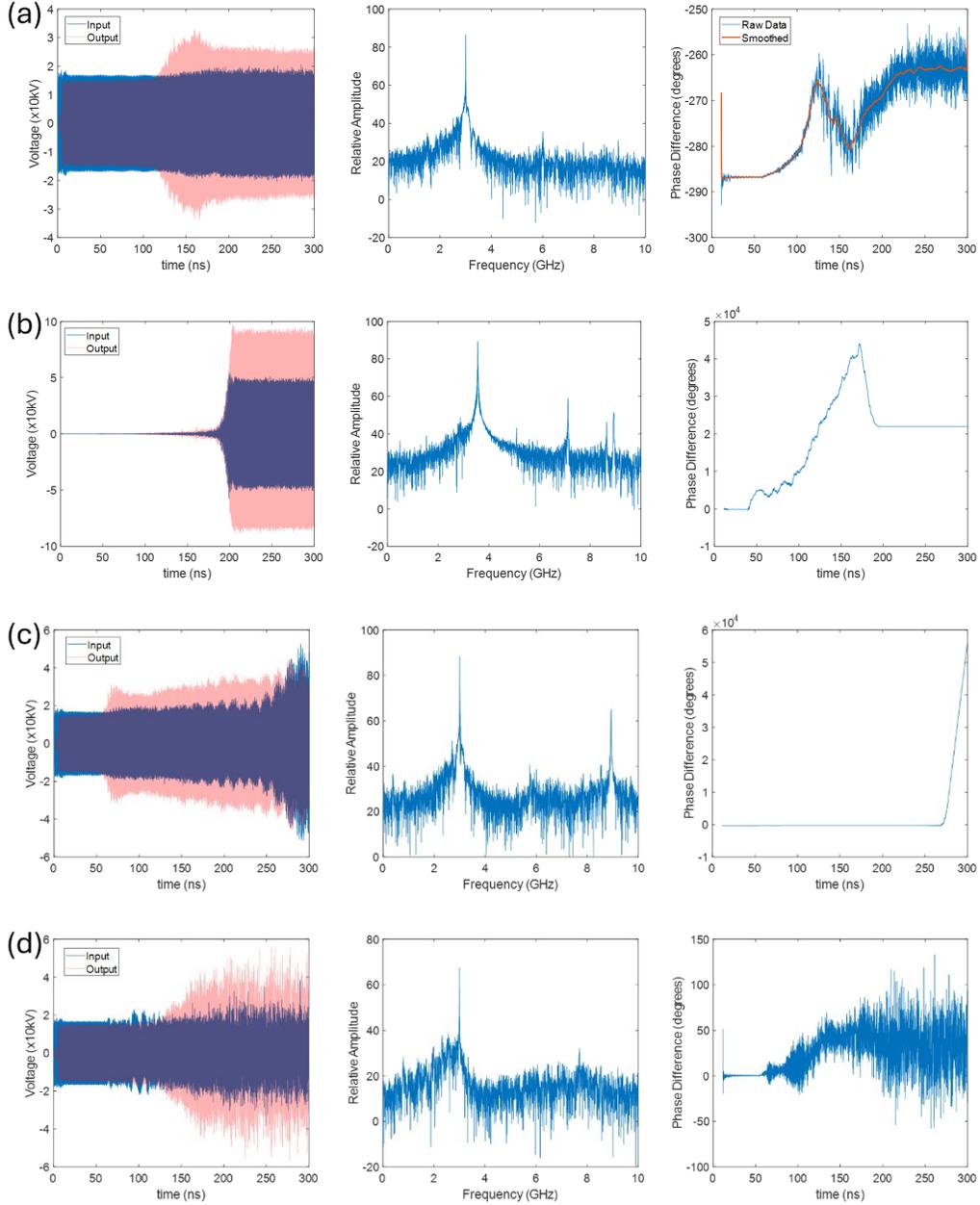

Figure 17: CST simulations of the RPCFA showing the input/output RF voltage traces with their respective frequency spectrum and phase difference for (a) low gain amplification operation at the nominal parameters with a smoothed phase difference and 3 GHz, 2 MW RF input, (b) strong oscillations in the 12 mm AK gap spacing geometry with zero-drive, (c) 20 mm AK gap spacing with high cathode current depicting the late time oscillations and 3 GHz, 2 MW RF input, and (d) 15 mm AK gap spacing and 3 GHz, 2 MW RF input with large variations in phase difference. Note the difference in scales for the phase difference plots.

The next feature of the phase difference in Fig. 17a is the dip (120 ns – 220 ns) which corresponds to the peak output power. From 120 ns to the end of the simulation, the phase difference curve is "noisy", there are high frequency jumps in phase that are $\pm 5$ degrees. This noise is a result of the time varying output phase, $\theta_o(t)$. Smoothing (red curve in Fig. 17a) can remove the noise and the impact of $\theta_o(t)$ on the phase difference. Eq. (4) predicts a frequency difference of 2 MHz for the linear trend in the dip. These small deviations in frequency calculated through the phase difference are exceedingly difficult to obtain directly through frequency analysis (FFT) where a very high time resolution in simulation or an extremely high sample rate in experiments would be required to discriminate the input signal and the frequency shifted signal. The final feature of the phase difference graph shows the steady state operation (220 ns – 300 ns) where the phase difference is relatively constant within ± 5 degrees and the frequency is 3 GHz. The ± 5 degrees of phase difference at steady state has been observed in other CFAs and has been reported by Gilmour [3]. For the RPCFA, the time dependence of the output phase can be described as $\theta_o(t) - \theta_i = \pm$ 5 degrees during steady state amplification.

The effects of strong self-excitations on the phase difference can be seen in Fig. 17b which has no RF input (zero drive). The noise-induced "input" could be reflections or backwards wave modes which will be uncorrelated to the output. The uncorrelated phase difference grows rapidly while the oscillator is evolving. The jagged shape in the phase difference plot depicts mode competition within the RPCFA; RF signals are mixture of all the different modes which can change the slope of the phase difference curve depending on the proportional strength of each mode within the total RF signal. Eventually, the device fully transitions to a steady state, standing wave 3.5 GHz oscillator and the phase difference becomes constant. The "input" and output become the same mode and thus fully correlated. This strong 3.5 GHz π-mode oscillation near Hull cutoff resembles MILO operation.

Both Figs. 17c and 17d show a deviation from nominal operation with Fig. 17c having late transition that occurs while the RPCFA is at steady state voltage. The phase difference plot clearly depicts the time at which the device oscillates. Fig. 17c is a current driven simulation where the cathode voltage was 600 kV and the current was 3 kA (seen in Fig. 6d). The device starts up faster due to the increased voltage and appears identical to the nominal case (a peak in output power before a dip to steady state); however, late in time (around 250 ns), the power begins to increase with time until the device oscillates. A 5.7 GHz frequency shift is predicted by Eq. (4) and is also seen in the FFT of the output signal; this high frequency mode is a higher order π/2-mode which has zero group velocity (not shown) and requires a higher velocity beam (>0.45c) for synchronization, so it is only seen in the high voltage/current simulation.

Figure 17d is a simulation with a 15 mm AK gap spacing which is the simulated transition point from zero-drive stable to zero-drive unstable (cf. Fig. 8). The frequency spectrum for this simulation looks similar to Fig. 17a, but with a slightly higher noise floor; the RF peak is 30 dB above noise floor compared to Fig. 17a which is 40 dB above the noise floor. However, the phase difference is much noisier (±50 degrees). The large high frequency phase difference corresponds to changes in harmonic content in the output signal. This suggests that the RPCFA is experiencing low power oscillations which affect the output signal.

From these simulations, the RPCFA can be seen to operate in three different states: amplifier (Fig. 17a), driven oscillator (Fig. 17d), and self-excited oscillator (Fig. 17b,c). By using the power, frequency, and phase difference plots, the operating state of the RPCFA and the transition of operating states can be determined.

**C. Experimental Phase Analysis**

Measurements of the phase difference from the RPCFA experiment can be used to determine whether the amplifier is oscillating for each shot. Figure 18 shows two shots with two different RF output power traces, but the frequency spectra are monochromatic with different noise levels. All shots in this experimental series exhibit the linear increase in the phase difference between the input and output RF signals (650 ns – 900 ns for Fig. 18a, 650 – 800 ns for Fig. 18b). This phase difference corresponds to a frequency shift of ~1-3 MHz, like what is seen in the simulations (Fig.17a). Late in time (e.g. 900 ns for Fig. 18a), the output signal is terminated either due to RF breakdown or plasma diode closure which reflects the signal, thus the phase difference late in time becomes erratic and unreliable. Therefore, whenever the output power falls to 0, the phase analysis is invalid. When the gain of the RPCFA is low, as seen in Fig. 18a, there are few deviations from the linear increase in phase during the voltage pulse except the relatively small 10 degree jump at peak output power (850 ns). The frequency spectrum has a lower noise floor as well. Overall, this experimental result agrees with the simulations, in that low gain amplifier shots show no signs of significant oscillations but have a noisy output phase.

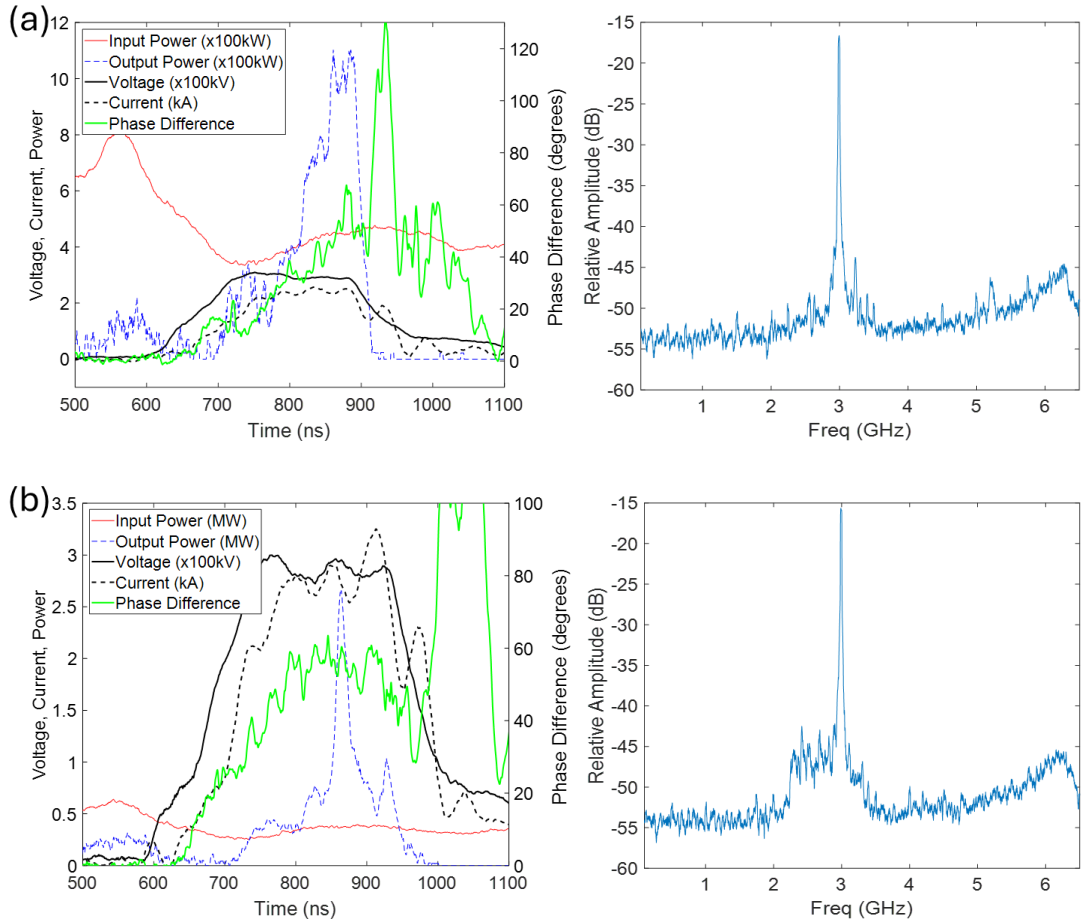

Figure 18: RPCFA experimental shots plotting: (left) the voltage, current, RF power, and phase difference and (right) the frequency spectra for (a) shot 19310 (~ 5 dB peak gain) with a 20 mm AK gap spacing and 420 kW RF input power and (b) shot 19055 (~ 9 dB) with an 18 mm AK gap spacing and 350 kW RF input power. Both AK gap spacings are zero-drive stable and were above the simulated threshold gap spacing for oscillations

With slightly higher gain, seen in Fig. 18b, the noise floor (2.4 GHz – 3.2 GHz) is increased substantially compared to Fig. 18a, almost 10 dB, yet the phase difference is still consistent with amplifier operation. With a stronger electron beam – RF wave interaction, there is more electronic noise imparted from the beam onto the signal to prime competing modes, but this does not imply that strong oscillations are present within the amplifier.

Another shot with a 20 mm AK gap (Fig. 19) depicts a completely different operation when compared to Fig. 18a, exhibiting higher output power and noise floor along with spikes in the phase difference (at 880 ns and 920 ns). The frequency spectrum appears similar to Fig. 18b with an increased noise floor compared to low gain amplifier shots. However, the spikes in the phase of 50-100° coincide with spikes in output power which imply oscillations during those high-power bursts. As seen before, the higher output power in the RPCFA leads to an increased noise floor, but this alone does not signify that strong oscillations are dominating the operation of the amplifier. This experimental shot is akin to the simulation in Fig. 17d which had an AK gap of 15 mm and was in the transition regime from amplifier to oscillator. Both exhibit a higher noise floor in the frequency spectrum and large jumps in phase difference. The RPCFA is acting as a hybrid amplifier/oscillator. If the gain of the device continues to increase, however, the amplifier transitions into a driven oscillator and the output power increases with the growth of the oscillations. From Fig. 19, the oscillation power growth was limited by the voltage pulse duration. Note that plasma diode closure effectively decreases the AK gap to the unstable range.

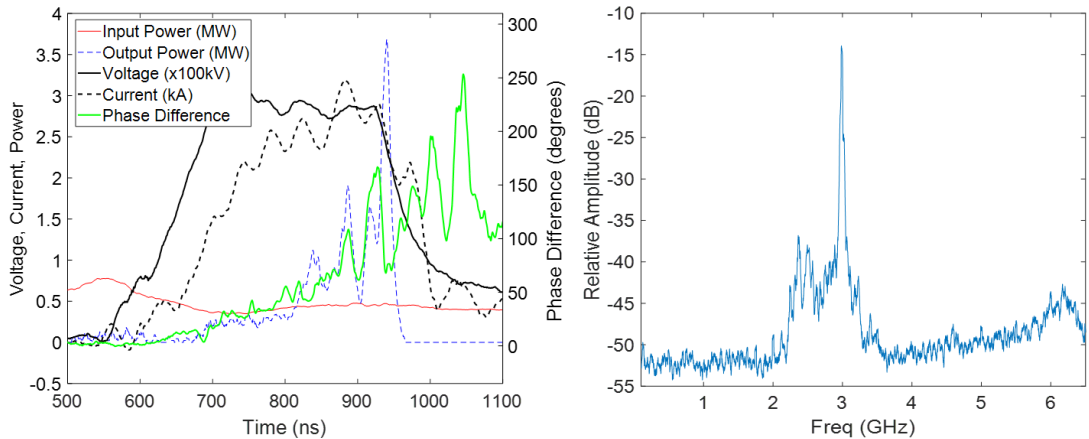

Figure19: RPCFA experimental shot plotting voltage, current, RF power, and phase difference, and (right) the frequency spectrum for shot 19311 with a 20 mm AK gap spacing and 440 kW RF input power. The 20 mm AK gap spacing is zero-drive stable and above the simulated threshold gap spacing for oscillations. Clear spikes in the phase difference mimic the spikes in output power denoting oscillations.

From previous experiments and simulations, the RPCFA is zero-drive unstable when the AK gap spacing is reduced to 12.5 mm. Figure 20 represents an experimental shot under those conditions with an RF input of 290 kW, where oscillations are expected. The frequency spectrum is similar to other high power shots; the noise floor (2.4 GHz – 3.2 GHz) is high compared to low gain shots (Fig. 18a). Output power shows some initial low gain steady state during the voltage rise (710 ns – 770 ns) and then a large spike in power which corresponds to a large spike in the phase difference (780 ns). The transition to oscillation (780 ns – 790 ns) when compared to the previous shot indicates that the growth rate of the oscillating mode is dependent on gap spacing as predicted by simulations. In both oscillating cases seen in Figures 19 and 20, there is not a steady state output power; the output power spikes and falls repeatedly. This phenomenon can be seen in the simulation (Fig. 17c after 250 ns) when the device is transitioning to an oscillator late in time. The dramatic swings in output power suggest that the device is transitioning between amplifying and oscillating where the conditions for steady state oscillation are not quite met. In all cases, the state of the RPCFA can only be determined by examining the frequency spectra, output power trace, and phase difference analysis. With the Hilbert transform phase difference analysis, the precise time of transition can be identified.

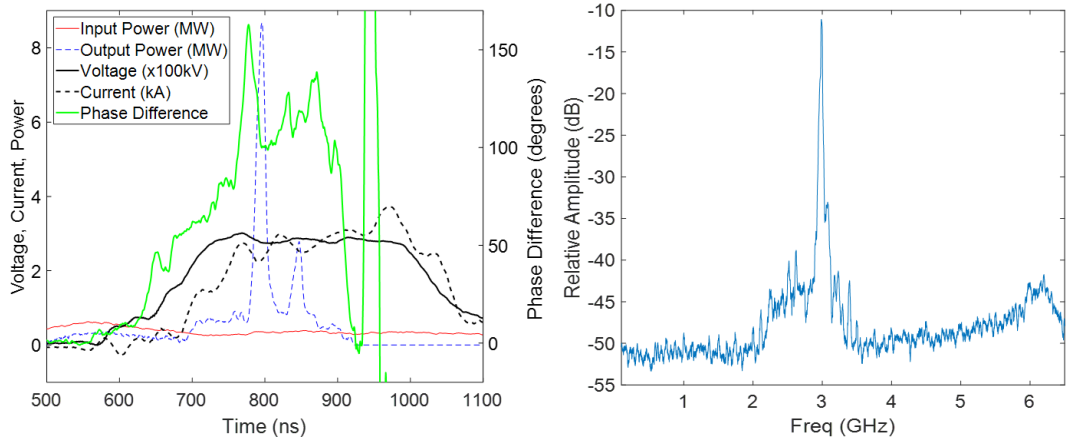

Figure 20: RPCFA experimental shot plotting voltage, current, RF power, and phase difference, and (right) the frequency spectrum for shot 19155 with a 12.5 mm AK gap spacing and 290 kW RF input power. The 12.5 mm AK gap spacing is not zero-drive stable and below the simulated threshold gap spacing for oscillations.

## V. Conclusion

In this paper, we examined the stability and characterized the operational modes of distributed emission crossed-field amplifiers via the RPCFA. A CFA can operate as an amplifier, driven oscillator, or a self-excited oscillator. In the nominal parameters (20 mm AK gap spacing), the RPCFA exhibited zero-drive stability and acted as a low gain amplifier with a linear growth regime. However, the operational mode of the RPCFA can be changed by a reduction in the AK gap spacing or by an increase in the driving current (cathode current). In both simulation and experiments, the transition from amplification to oscillation was measured and was shown to occur around the same degree of magnetic insulation. The oscillation frequency changed depending on the AK gap spacing which led to a change in the operating mode, as shown in the cold characteristics of the RPCFA. With reductions in the AK gap spacing, the Q factor for each mode can change significantly and independently despite little change in the dispersion relation which can lead to excitation of undesired modes. The RF circuit might interact with electrons in the beam with a much lower velocity than the velocity at the top of the Brillouin hub, leading to undesirable mode competition and noise. When the AK gap spacing is sufficiently small, the CFA operates close to Hull cutoff (like a MILO) and oscillations are observed.

This paper showed that an examination of the output power trace and the frequency spectrum is not enough to determine the exact mode of operation. By including an instantaneous phase analysis by Hilbert transforms, the operational mode of the RPCFA can be determined as well as the timing for any transition between operational modes. A constant slope in a phase difference vs time indicates a change in frequency between the input and output. Large changes in the phase relation are a sign of growing oscillations in the CFA and an examination of the frequency spectra can be used to determine whether the oscillation was driven or self-excited. The methods and analysis in this paper can be used to improve CFA design and diagnostics. The Hilbert transform may similarly yield new insights into simulations and experiments on other types of amplifiers, and on injection locking.

## Acknowledgements

This research was supported by the Air Force Office of Scientific Research (AFOSR) Grant No. FA9550 20-1-0409 through a Multidisciplinary University Initiative (MURI) managed by the University of New Mexico, by the AFOSR Grant No. FA9550 21-1-0184, and by the Office of Naval Research (ONR) Award No. N00014-23-1-2143. We would like to thank Dr. John W. Luginsland for his insightful ideas and for encouraging us to investigate the phase of the RPCFA.

## References

[1] G. B. Collins, *Microwave Magnetrons*, New York, NY, USA: McGraw-Hill MIT Radiation Laboratory Series, 1948.

[2] E. Okress, *Crossed-Field Microwave Devices*, New York, NY, USA: Academic Press, 1961.

[3] A. S. Gilmour, *Klystrons, Travelling Wave Tubes, Magnetrons, Crossed-Field Amplifiers, and Gyrotrons*, Artech House, 2011.

[4] P. H. Siegel, "Microwaves are everywhere: "ovens: from magnetrons to metamaterials"," IEEE Journal of Microwaves, vol. 1, no. 2, pp. 523-531, April 2021.

[5] R. R. Warnecke, W. Kleen, A. Lerbs, O. Dohler and H. Huber, "The magnetron-type traveling-wave amplifier tube," Proceedings of the IRE, vol. 38, no. 5, pp. 486-495, May 1950.

[6] A. W. Hull, "The effect of a uniform magnetic field on the motion of electrons between coaxial cylinders," Physical Review, vol. 18, no. 1, pp. 31-57, July 1921.

[7] W. Brown, "Description and operating characteristics of the platinotron - a new microwave tube device," in Proceedings of the IRE, vol. 45, no. 9, pp. 1209-1222, September 1957.

[8] T. E. Ruden, "The amplitron as a high power, efficient, RF power source for long pulse, high resolution linear accelerators," IEEE Transactions on Nuclear Science, vol. 12, no. 3, pp. 169-173, June 1965.

[9] J. F. Skowron, "The continuous-cathode (emitting-sole) crossed-field amplifier," Proceedings of the IEEE, vol. 61, no. 3, pp. 330-356, March 1973.

[10] R. G. Carter, *Microwave and RF Vacuum Electronic Power Sources*, Cambridge University Press, 2018.

[11] D. Chen, "A study of the linear emitting-sole magnetron amplifier," Ph.D. dissertation, Standford University, Palo Alto, CA, USA 1959.

[12] G. K. Farney, "CFA design improvement program volume I - Instrumented CFA studies," Varian Associates, Inc., June 1978. https://apps.dtic.mil/sti/pdfs/ADA065735.pdf

[13] H. L. McDowell, "CFA design improvement program volume II - Computer modeling studies," Varian Associates, Inc., June 1978. https://apps.dtic.mil/sti/pdfs/ADA065736.pdf

[14] M. C. Balk, "Simulation possibilities of vacuum electronic devices with CST PARTICLE STUDIO$^{TM}$," in *IEEE International Vacuum Electronics Conference* (IEEE 2008), pp. 459-460, Monterey, CA, 2008.

[15] S. C. Exelby, G. B. Greening, N. M. Jordan, D. A. Packard, D. Simon, Y. Y. Lau, B. W. Hoff and R. M. Gilgenbach, "High-power recirculating planar crossed-field amplifier design and development," IEEE Transactions on Electron Devices, vol. 65, no. 6, pp. 2361-2365, June 2018.

[16] S. C. Exelby, G. B. Greening, N. M. Jordan, D. A. Packard, D. Simon, Y. Y. Lau, B. W. Hoff and R. M. Gilgenbach, "High-power amplification experiments on a recirculating planar crossed-field amplifier," IEEE Transactions on Plasma Science, vol. 48, no. 6, pp. 1917-1922, June 2020.

[17] M. Pearlman, D. N. Smithe, M. Worthington, J. Watrous, A. Garner and J. J. Browning, "Spoke characterization in re-entract backward wave crossed-field amplifiers via simulation," IEEE Transactions on Electron Devices, vol. 71, no. 8, pp. 5020-5027, August 2024.

[18] M. Pearlman, D. N. Smithe, C. Roark, M. Worthington, J. Watrous, A. Garner and J. J. Browning, "Simulation of a pulsed 4.7 MW L-band crossed-field amplifier," IEEE Transactions on Electron Devices, vol. 69, no. 12, pp. 7053-7058, 2022.

[19] R. MacGregor, C. Chan, J. Ye and T. Ruden, "A circular crossed-field amplifier for in situ measurements, study of reentrant beam effects, and comparison with numerical simulation," IEEE Transactions on Electron Devices, vol. 41, no. 8, pp. 1456-1464, August 1994.

[20] D. Chernin, A. Drobot, G. Hilfer, M. Kress and S. Riyopoulos, "Computer studies of noise generation in crossed-field amplifiers," in *IEEE International Electron Devices Meeting* (IEEE 1991), pp. 593-596, Washington DC, USA, 1991.

[21] S. Riyopoulos, "Feedback-induced noise in crossed-field devices," IEEE Transactions on Plasma Science, vol. 20, no. 3, pp. 360-369, June 1992.

[22] D. Chernin, "Computer simulations of low noise states in a high-power crossed-field amplifier," IEEE Trans. Electron Devices, vol. 43, no. 11, pp. 2004-2010, November 1996

[23] G. E. Dombrowsky, "Simulation of magnetrons and crossed-field amplifiers," IEEE Trans. Electron Devices, vol. 35, no. 11, pp. 2060-2067, November 1968.

[24] H. McDowell, "Computer modeling of CFA's using moving wavelength and trajectory tracing codes," in 1994 Microwave Power Tube Conference, Monterey, CA, paper 2D.4.

[25] R. M. Gilgenbach, Y. Y. Lau, H. McDowell, K. L. Cartwright, and T. A. Spencer, "Crossed-field devices," in *Modern Microwave and Millimeter Wave Power Electronics*, edited by R. J. Barker, N. C. Luhmann, J. H. Booske, and G. S. Nusinovich, IEEE, Piscataway, NJ, 2004 Chap. 6.

[26] J. C. Slater, *Microwave Electronics*, New York, NY: Van Nostrand, 1951.

[27] L. Brillouin, "Theory of the magnetron. I," Physical Review, vol. 60, no. 5, pp. 385-396, September 1941.

[28] O. Buneman, "Symmetrical states and their breakup," in *Crossed-Field Microwave Devices*, New York, NY, USA, Academic Press, 1961, pp. 209-233.

[29] A. Palevsky, G. Bekefi, and A. T. Drobot, "Numerical simulation of oscillating magnetrons," Journal of Applied Physics, Vol. 52, no. 8, pp. 4938–4941, August 1981.

[30] P. J. Christenson, D. P. Chernin, A. L. Garner and Y. Y. Lau, "Resistive destabilization of cycloidal electron flow and universality of (near-) Brillouin flow in a crossed-field gap," Physics of Plasmas, vol. 3, no. 12, pp. 4455-4462, December 1996.

[31] Y. Y. Lau, D. A. Packard, C. J. Swenson, J. W. Luginsland, D. Li, A. Jassem, N. M. Jordan, R. D. McBride and R. M. Gilgenbach, "Explicit Brillouin flow solutions in magnetrons, magnetically insulated line oscillators, and radial magnetically insulated transmission lines," IEEE Transactions on Plasma Science, vol. 49, no. 11, pp. 3418-3437, November 2021.

[32] M. Lopez, Y. Y. Lau, J. W. Luginsland, D. W. Jordan and R. M. Gilgenbach, "Limiting current in a relativistic diode under the condition of magnetic insulation," Physics of Plasmas, vol. 10, no. 11, pp. 4489-4493, November 2003.

[33] R. M. Gilgenbach, Y. Y. Lau, D. M. French, B. W. Hoff, J. Luginsland and M. Franzi, "Crossed field device". United States Patent 8,841,867, September 2014.

[34] B. W. Hoff, S. Beeson, D. Simon, W. Tang, R. Smith, S. C. Exelby, N. M. Jordan, A. Savir, R. M. Gilgenbach, P. D. Lepell and T. Montoya, "Brazed carbon fiber fabric field emission cathode," Review of Scientific Instruments, 064702, vol. 91, no. 6, June 2020.

[35] R. Revolinsky, C. Swenson, N. Jordan, Y. Y. Lau, R. M. Gilgenbach, "On the two-dimensional Brillouin flow," Physics of Plasmas, 053109, vol. 31, no. 5, May 2024.

[36] P. J. Christenson, Y. Y. Lau, "Transition to turbulence in a crossed-field gap," Physics of Plasmas, vol. 1, no. 12, pp. 3725-3727, December 1994, Erratum vol. 3, no. 11, pp. 4293, November 1996.

[37] D. Chernin, A. Jassem, Y. Y. Lau, "Thermal electron flow in a planar crossed-field diode", IEEE Transactions on Plasma Science, vol. 48, no. 9, pp. 3109-3114, September 2020.

[38] D. A. Packard, Y. Y. Lau, E. N. Guerin, C. J. Swenson, S. V. Langellotti, A. Jassem, D. Li, N. M. Jordan, J. W. Luginsland, R. D. McBride, R. M. Gilgenbach, "Theory, simulation, and experiments on a magnetically insulated line oscillator (MILO) at 10 kA, 240 kV near Hull cutoff condition," Physics of Plasmas, vol. 28, no. 12, 123102, November 2021.

[39] F. Antoulinakis, P. Wong, A. Jassem and Y. Y. Lau, "Absolute instability and transient growth near the band edges of a traveling wave tube," Physics of Plasmas, 072102, vol. 25, no. 7, June 2018.

[40] R. M. Gilgenbach, L. D. Horton, R. F. Lucey, S. Bidwell, M. Cuneo, J. Miller, and L. Smutek, "Microsecond electron beam diode closure experiments," in Invited Talk, Proceedings of the IEEE Pulsed Power Conference, Crystal City, VA, 1985, p. 7.

[41] H. E. Rowe, *Signals and Noise in Communication Systems*, Princeton, New Jersey: D. Van Nostrand Company. Inc., 1965.

[42] A. H. McCurdy, A. K. Ganguly and C. M. Armstrong, "Operation and theory of a driven single-mode electron cyclotron maser," Physical Review A, vol. 40, no. 3, pp. 1402-1421, August 1989.